\documentclass[twocolumn,aps,superscriptaddress,prb]{revtex4-1}
\setcitestyle{numbers,square}
\usepackage{amsmath,amssymb}
\usepackage{hyperref}
\usepackage{tabularx}
\usepackage{graphicx}
\usepackage{bmpsize}
\usepackage{bm}
\usepackage{color}
\usepackage{amssymb}
\usepackage[version=3]{mhchem}
\usepackage{newtxmath}
\usepackage{braket}
\usepackage{comment}
\usepackage[normalem]{ulem} 
\usepackage{float}
\DeclareMathAlphabet{\mathpzc}{OT1}{pzc}{m}{it}
\newcommand{\nn}{\nonumber}

\usepackage{mathtools}
\usepackage{physics}

\newcommand{\blue}[1]{\textcolor{black}{ #1}}

\newcommand{\kawa}[1]{\textcolor{black}{ #1}}

\begin{document}
\title{One-dimensional moir\'e engineering in zigzag graphene nanoribbons on hBN}
\author{Ryosuke Okumura}
\affiliation{Department of Physics, Osaka University, Toyonaka, Osaka 560-0043, Japan}
\author{Naoto Nakatsuji}
\affiliation{Department of Physics and Astronomy, Stony Brook University, Stony Brook, NY, 11794, USA}
\affiliation{Department of Physics, Osaka University, Toyonaka, Osaka 560-0043, Japan}
\author{Takuto Kawakami}
\affiliation{Department of Physics, Osaka University, Toyonaka, Osaka 560-0043, Japan}
\author{Mikito Koshino}
\affiliation{Department of Physics, Osaka University, Toyonaka, Osaka 560-0043, Japan}
\date{\today}

\begin{abstract}
We study the structural relaxation and electronic properties of a one-dimensional (1D) moir\'e system composed of a zigzag graphene nanoribbon (GNR) placed on a hexagonal boron nitride (hBN) substrate. Using an effective grid model derived from continuum elasticity theory, we calculate the relaxed atomic structure of the GNR/hBN system for various twist angles and ribbon widths. The relaxation gives rise to a characteristic 1D domain structure consisting of alternating commensurate AB$'$ regions and two distinct types of domain boundaries. 
At finite twist angles, the ribbon adopts a wavy shape, locally tracing the hBN zigzag direction but occasionally sliding to adjacent atomic rows.
The resulting moir\'e potential strongly modulates the electronic structure: the zero-energy zigzag edge states are modulated by the local stacking, leading to densely packed subbands in the AB$'$ domains and sharply localized domain-wall states in the energy gaps between domain plateaus, which together realize gate-tunable one-dimensional arrays of quantum-confined electronic states.
Our results demonstrate that moir\'e modulation in GNR/hBN heterostructures provides a versatile platform for electronic structure engineering and the design of 1D moir\'e nanodevices.
\end{abstract}

\maketitle

\section{introduction}
\label{sec:intro}

In recent years, remarkable progress has been made in the study of two-dimensional (2D) moir\'e materials, triggered by the discovery of various exotic phenomena in twisted bilayer systems. The electronic structures of these materials are strongly modulated by the long-range periodic potential originating from the moir\'e superlattice
\cite{bistritzer2011moirepnas,PhysRevB.84.035440,PhysRevB.86.155449,cao2018_80,cao2018_43,doi:10.1126/science.aav1910,Kerelsky2019,xie2019spectroscopic,jiang2019charge,polshyn2019large,Choi2019,doi:10.1126/science.aaw3780,lu2019superconductors,PhysRevLett.124.076801,doi:10.1126/science.aay5533,chen2020tunable,saito2020independent,zondiner2020cascade,wong2020cascade,stepanov2020untying,arora2020superconductivity,PhysRevLett.127.197701}.
Such moir\'e-induced electronic reconstruction has provided a versatile platform for exploring correlated and topological quantum phenomena in van der Waals heterostructures\cite{doi:10.1021/acs.nanolett.8b05061,finney2019tunable,PhysRevB.104.035306,PhysRevResearch.4.013028,PhysRevLett.121.026402,li2021continuous,ghiotto2021quantum,li2021quantum,PhysRevB.104.075150,PhysRevB.111.125410}.



While most previous studies have focused on 2D moir\'e systems formed by lattice mismatch or rotational misalignment between two 2D layers, recent work has begun exploring hybrid moir\'e systems that combine one-dimensional (1D) and 2D materials.
Experimentally, catalytic growth of graphene nanoribbons (GNRs) on hBN has revealed clear 1D moir\'e patterns reflecting the atomic registry of the substrate \cite{lyu2022catalytic,lyu2024graphene}.
Theoretically, numerical studies have examined the interfacial mechanics of GNRs on hBN, highlighting the interplay between in-plane elasticity and interfacial registry \cite{ouyang2018nanoserpents,xue2022peeling}.
For the electronic properties of GNR/hBN systems, band-structure calculations have been performed for perfectly commensurate geometries without moir\'e patterns \cite{gani2018electronic}.
The moir\'e effect on the electronic structure has been studied in other 1D systems, including carbon nanotubes on graphene \cite{PhysRevResearch.2.022041} and hBN \cite{zhou2022moire}, double-wall carbon nanotubes \cite{liu2014van,koshino2015incommensurate,zhao2020observation,zhao2022interlayer}, and collapsed chiral carbon nanotubes \cite{arroyo2020one}.

However, the structural relaxation in 1D–2D moir\'e systems and its impact on the electronic properties, including its dependence on the relative twist angle and ribbon width, remain largely unexplored.
Structural relaxation generally plays a crucial role in determining the physical properties of moir\'e materials \cite{brown2012twinning,PhysRevB.101.224107,shin2021electron,PhysRevB.98.224102,PhysRevLett.124.206101,xue2011scanning,doi:10.1021/nl2005115,woods2014commensurate,PhysRevB.90.075428,jung2015origin,mcgilly2020visualization,krisna2023moire,doi:10.1021/acs.nanolett.4c04201,doi:10.7566/JPSJ.94.044602}. In 2D graphene/hBN moir\'e superlattices, for example, the relaxation forms a 2D periodic lattice of commensurate AB$'$ domains, in which the carbon atoms of graphene are aligned vertically above boron atoms of hBN, separated by narrow domain walls \cite{xue2011scanning,doi:10.1021/nl2005115,woods2014commensurate,PhysRevB.90.075428,jung2015origin,mcgilly2020visualization,krisna2023moire,doi:10.1021/acs.nanolett.4c04201,doi:10.7566/JPSJ.94.044602}.

Here, we calculate the relaxed atomic structure of the GNR/hBN system for various twist angles using an effective grid model, a discretized formulation of the continuum elasticity theory \cite{nam2017lattice,krisna2023moire}.
In the absence of relaxation, the moir\'e pattern of a GNR on hBN corresponds to a partial segment of the 2D graphene/hBN moir\'e superlattice. Upon relaxation, however, the system exhibits a peculiar 1D moir\'e pattern that does not coincide with any portion of the relaxed 2D moir\'e structure.

At zero twist angle, the system forms a serial array of AB$'$ domains separated by uniform domain walls along the ribbon, resulting in a fully 1D configuration. As the twist angle increases, the GNR develops a wavy geometry, where the ribbon locally follows the zigzag orientation of the hBN lattice but occasionally shifts laterally to adjacent atomic lanes to accommodate the rotation. The resulting pattern consists of 1D sequences of domains separated by two distinct types of domain walls, referred to as the $\alpha$ and $\beta$ structures, corresponding respectively to relative atomic shifts along the ribbon axis and in the perpendicular direction.

The electronic properties are investigated within a tight-binding framework. We find that the zero-energy zigzag edge states of graphene \cite{fujita1996peculiar, nakada1996edge, son2006energy, son2006half}  are strongly modulated by the moir\'e potential from hBN. The local density of states (LDOS) closely follows the effective potential arising from interlayer coupling: the potential is nearly constant within AB$'$ domains, giving rise to densely distributed subbands localized in those regions. The top and bottom edges experience markedly different potential energies (by about 40 meV) due to their distinct atomic alignments with respect to hBN, leading to a clear energy separation between the corresponding edge states. At the domain boundaries, the potential exhibits sharp peaks, resulting in sparsely distributed states located between the upper and lower domain bands. 
Each of these domain-wall states is strongly localized with a spatial extent of only a few atomic lattice constants,
thereby realizing one-dimensional arrays of quantum-confined electronic states.

These findings demonstrate that 1D/2D moir\'e systems, exemplified by GNR/hBN heterostructures, exhibit unique relaxation patterns and moir\'e-modulated edge physics that are fundamentally distinct from those in conventional 2D moir\'e materials. 
Such one-dimensional moir\'e architectures may provide new design principles for nanodevices, where both the structural and electronic degrees of freedom can be engineered through controlled twisting and substrate selection.

The paper is organized as follows.
In Sec.~\ref{sec_methods}, we introduce the effective grid model derived from the continuum elasticity approach.
In Sec.~\ref{sec:OLS}, we present the lattice relaxation of GNR/hBN systems for various twist angles and ribbon widths.
In Sec.~\ref{sec:ES}, we calculate the electronic band structures and local density of states using the tight-binding model.
A brief conclusion is given in Sec.~\ref{sec:conclusion}.


\section{Effective theory for structural relaxation in GNR/\lowercase{h}BN}

\label{sec_methods}

\subsection{Superlattice geometry of GNR/hBN}
\label{sec_geometry}

\begin{figure}
        \begin{center}
        \leavevmode\includegraphics[width=1. \hsize]
        {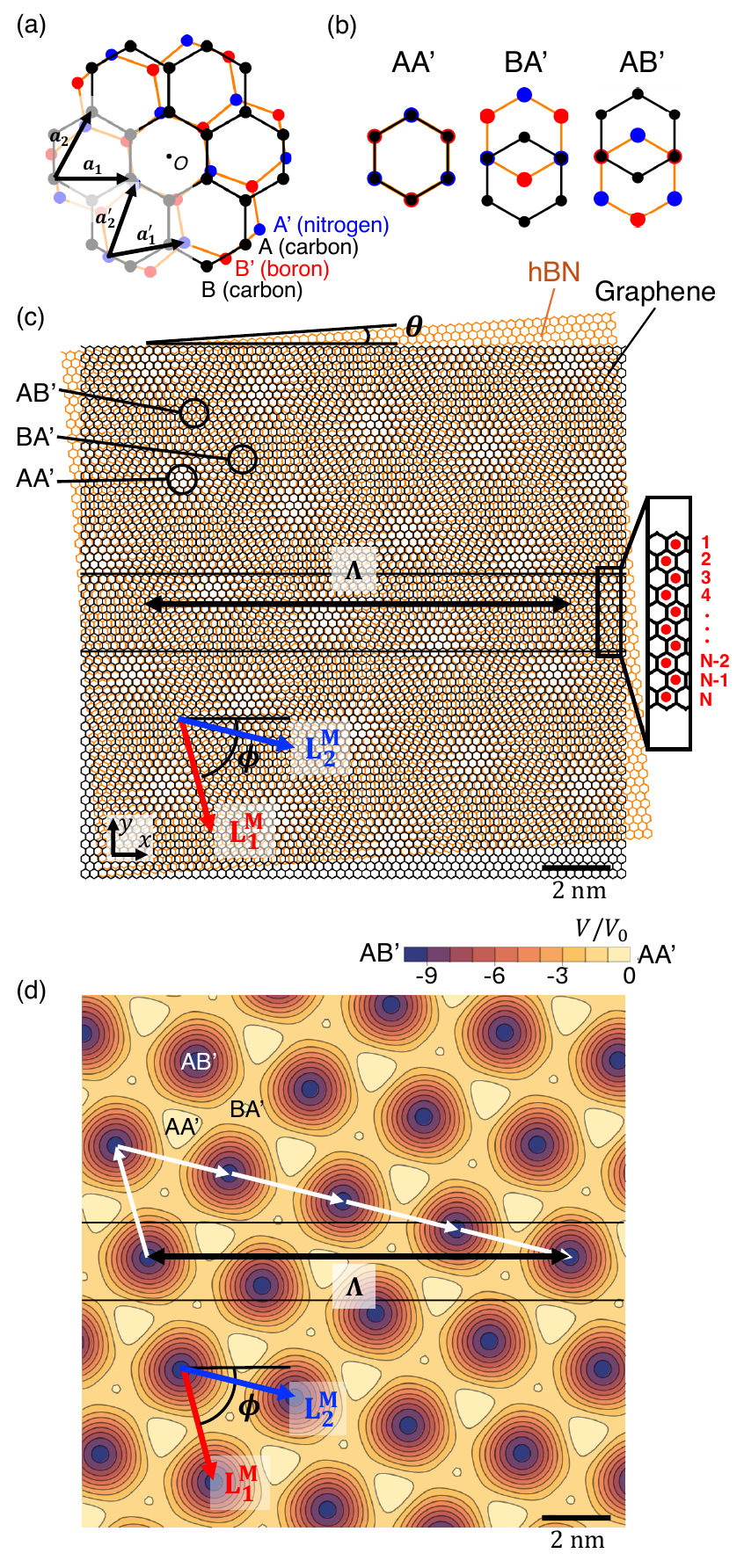}
     \caption{
Schematic of the graphene/hBN moir\'e superlattice. (a) Atomic configuration. (b) Local stacking arrangements: AA$'$, AB$'$, and BA$'$. (c) Lattice structure and moir\'e pattern at $\theta = 4.08^\circ$ with $(m,n)=(-1,4)$. The horizontal lines indicate the GNR region with $N=10$. (d) Contour plot of the interlayer binding energy corresponding to panel (c). The one-dimensional superlattice period is given by a linear combination of the moir\'e lattice vectors, $\mathbf{\Lambda} = m \mathbf{L}^{\mathrm{M}}_1 + n \mathbf{L}^{\mathrm{M}}_2$, represented by the white arrows. 
        }
        \label{g_hbn_4.08deg}
        \end{center}
    \end{figure}

We consider a zigzag graphene nanoribbon (GNR) placed on a hexagonal boron nitride (hBN) substrate.
Both graphene and hBN have honeycomb lattices with  lattice constants
$a \simeq 0.246$ nm and $a' \simeq 0.2505$ nm, respectively.
We define graphene sublattices A, B, and hBN sublattices A$'$ (nitrogen), B$'$ (boron) as in Fig.~\ref{g_hbn_4.08deg}(a).
We construct the GNR/hBN system by starting from a two-dimensional (2D) moir\'e superlattice of graphene on hBN with twist angle $\theta$, and then cutting a GNR parallel to the zigzag direction.
The twist angle $\theta$ is defined as the relative orientation angle of the hBN lattice with respect to that of graphene, measured from the configuration where the two honeycomb lattices are parallel.
Figure~\ref{g_hbn_4.08deg}(c) displays the full superlattice structure, where the horizontal lines mark the GNR region with $N=10$.
Here $N$ is the width of the zigzag GNR, which is defined by the number of hexagons across the ribbon as illustrated in the right panel of Fig.~\ref{g_hbn_4.08deg}(c).
Throughout this work, we take the $x$ axis to be aligned with the GNR.

\begin{figure}[h]
        \begin{center}
        \leavevmode\includegraphics[width=1. \hsize]
        {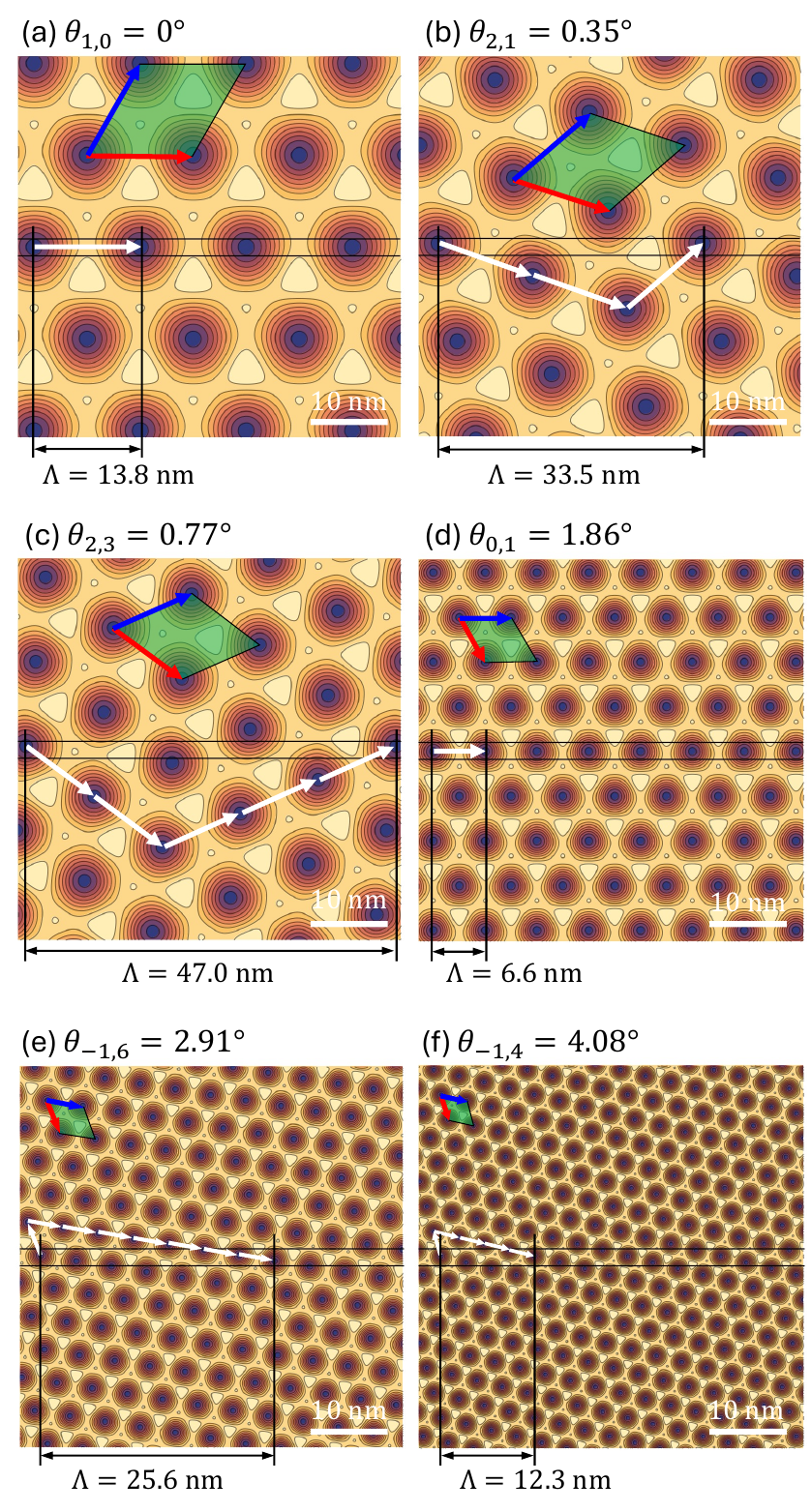}
     \caption{
     Moir\'e patterns for different twist angles, similar to Fig.~\ref{g_hbn_4.08deg}(d). In each panel, the green hexagon represents a unit cell of the 2D moir\'e pattern, spanned by the moir\'e lattice vectors $\mathbf{L}^\mathrm{M}_1$ (red arrow) and $\mathbf{L}^\mathrm{M}_2$ (blue arrow). The horizontal lines indicate the GNR region with $N = 10$.
        }
        \label{lambda_m}
        \end{center}
    \end{figure}

The moir\'e pattern of graphene/hBN system is characterized by period
\begin{equation}\label{eq:lm}
L_{\mathrm{M}}=\frac{(1+\epsilon)a}{\sqrt{\epsilon^{2}+2(1+\epsilon)(1-\cos{\theta})}},
\end{equation}
and relative angle,
\begin{equation}\label{phi}
\phi=\arctan\left(\frac{-\sin{\theta}}{1+\epsilon-\cos{\theta}}\right),
\end{equation}
where $\epsilon = a'/a - 1 \simeq 0.018$~\cite{moon2014electronic}.
The primitive lattice vectors of the moir\'e pattern are written as $\mathbf{L}^\mathrm{M}_1=L_{\mathrm{M}}\hat{\mathbf{e}}_{\phi}$ and
$\mathbf{L}^\mathrm{M}_2= L_{\mathrm{M}} \hat{\mathbf{e}}_{\phi + 60^\circ}$,
with $\hat{\mathbf{e}}_\phi=(\cos{\phi},\sin{\phi})$.
In Fig.~\ref{g_hbn_4.08deg}(d), we visualize the moir\'e pattern using the contrast of the local binding energy $V_{\rm B}$ (its precise definition is given in Sec.~\ref{sec_relaxation}).
The positions AA$'$, BA$'$, and AB$'$ correspond to the characteristic local stackings 
illustrated in Fig.~\ref{g_hbn_4.08deg}(b),
among which the AB$'$ stacking is the most stable.
Figure \ref{lambda_m} shows the moir\'e pattern with different twist angle $\theta$.
As $\theta$ increases from zero, the moir\'e periodicity decreases, while the moir\'e angle $\phi$ (the angle of $\mathbf{L}^\mathrm{M}_1$ with respect to the $x$ axis) rotates negatively from the initial value $\phi=0$.

For a general twist angle $\theta$, the GNR/hBN system is not periodic along the $x$ direction.
A periodicity arises, however, when a moir\'e lattice vector $\mathbf{L}(m,n)=m\mathbf{L}^{\mathrm{M}}_{1}+n\mathbf{L}^{\mathrm{M}}_{2}$ ($m, n$: integers)
is aligned with the GNR direction (the $x$ axis), in which case the one-dimensional superlattice period is given by $\mathbf{\Lambda}=\mathbf{L}(m,n)$.
We denote the twist angles satisfying this condition as $\theta=\theta_{m,n}$.
Several values of $\theta_{m,n}$ that yield relatively small 1D period are listed in Table~\ref{tab:parameters}.
The configuration shown in Fig.~\ref{g_hbn_4.08deg} corresponds to $(m,n)=(-1,4)$, for which the twist angle is $\theta =4.08^\circ$.


\begin{table}[h]
\caption{
Index $(m,n)$, twist angle $\theta_{m,n}$, moir\'e periods $\mathrm{L}^{\rm M}_i$ and 1D superlattice period $\Lambda$, and moir\'e orientation angle $\phi$ for the GNR/hBN systems considered in this study.
}
\label{tab:parameters}
\begin{tabular}{ccccc}
    \hline\hline
    $(m,n)$
    &$\theta_{m,n}$
    &$L_{\mathrm{M}}\ [\mathrm{nm}]$
    &$\mathrm{\Lambda}\ [\mathrm{nm}]$
    &$\phi$ \\
    \hline
    (1, 0)  & 0$^\circ$  &  13.8  & 13.8  & $0^\circ$ \\
    (2, 1) & 0.35$^\circ$  & 13.0 & 33.5  & $-18.6^\circ$ \\
    (2, 3) & 0.77$^\circ$  & 11.0  & 47.0 & $-36.3^\circ$ \\
    (0, 1) & 1.86$^\circ$  & 6.6 & 6.6 & $-60.0^\circ$ \\
    $(-1, 6)$ & 2.91$^\circ$ & 4.6  & 25.6 & $-69.0^\circ$ \\
    $(-1, 4)$ & 4.08$^\circ$ & 3.3  & 12.3 & $-73.8^\circ$ \\
    \hline\hline
\end{tabular}
\end{table}

\subsection{Effective model}
\label{sec_relaxation}

\begin{figure}[h]
    \begin{center}
    \leavevmode\includegraphics[width=1. \hsize]
    {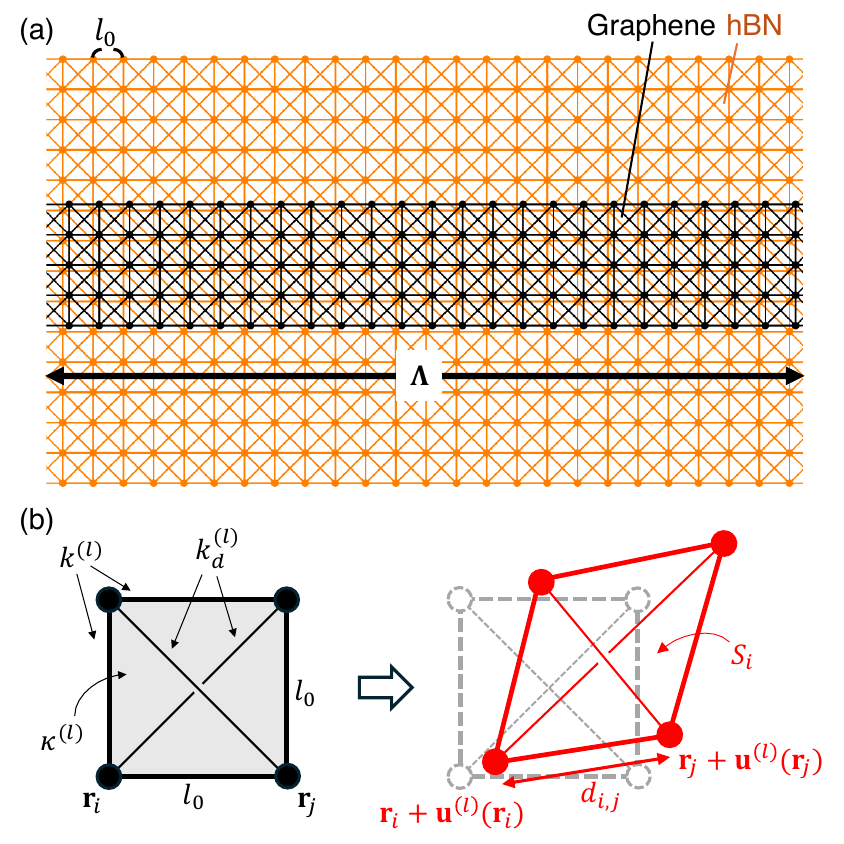}
    \caption{
    (a) Discrete square-grid model. Graphene and hBN layers are represented by parallel square grids with spacing $l_0$ and identical orientation, independent of the twist angle. Grid vertices correspond to mass points, and springs connect them along orthogonal and diagonal directions.
(b) A single grid cell of the spring–mass model and its deformation (see text).
}
    \label{grid}
    \end{center}
\end{figure}

We model the lattice relaxation in the GNR/hBN heterostructure using
an effective spring-mass approach, based on the continuum elasticity framework.
We first describe the continuum elasticity description that has been used
for lattice relaxation in graphene/hBN moir\'e systems~\cite{krisna2023moire, koshino2019moire, nam2017lattice}.
We consider the in-plane displacement $\mathbf{r}\rightarrow \mathbf{r}+\mathbf{u}^{(l)}(\mathbf{r})$, where $\mathbf{r}=(x,y)$ is the position, and $\mathbf{u}^{(l)}=(u_x^{(l)},u_y^{(l)})$ denotes displacement at $\mathbf{r}$ on layer $l$ ($l=1$ for graphene and $l=2$ for hBN).
The total energy in the presence of $\mathbf{u}^{(l)}$ is written as
\begin{equation}\label{eq:eplusb}
    U=U_{\mathrm{E}} + U_{\mathrm{B}}
\end{equation}
where $U_{\mathrm{E}}$ and $U_{\mathrm{B}}$ denote the elastic and interlayer binding energies, respectively, expressed as functionals of $\mathbf{u}^{(l)}(\mathbf{r})$.
The elastic energy takes the standard isotropic form~\cite{nam2017lattice, ando2002presence},
\begin{align}\label{eq:vecont}
    U_{\mathrm{E}} = & \sum_{l=1,2} \int  \frac{1}{2}\bigg( (\lambda^{(l)}+\mu^{(l)})(u_{xx}^{(l)}+u_{yy}^{(l)})^2 \nn\\
    &+\mu^{(l)}\big[(u_{xx}^{(l)}-u_{yy}^{(l)})^2+4(u_{xy}^{(l)})^2 \big] \bigg)d^2\mathbf{r}
\end{align}
with strain tensor components $u_{\alpha\beta}^{(l)}=\frac{1}{2}(\partial_{\alpha}u_\beta^{(l)}+\partial_{\beta}u_\alpha^{(l)})$, 
where $\alpha,\beta\in \{x,y\}$.
The Lam\'e parameters are chosen as
$\lambda^{(1)}=3.25~\mathrm{eV/\mathring{A}^2}$, $\mu^{(1)}=9.57~\mathrm{eV/\mathring{A}^2}$ for graphene,
and $\lambda^{(2)}=3.5~\mathrm{eV/\mathring{A}^2}$, $\mu^{(2)}=7.8~\mathrm{eV/\mathring{A}^2}$ for hBN~\cite{jung2015origin, zakharchenko2009finite, sachs2011adhesion}.
The interlayer binding energy in the continuum description is given by
\begin{align}\label{eq:vbcont}
&U_{\mathrm{B}} = \int V_{\mathbf{B}}(\mathbf{r})\mathrm{d}^2 \mathbf{r},
\end{align}
and, for systems with sufficiently long moir\'e periods,
the local binding energy density can be written as~\cite{krisna2023moire}
\begin{align}
    \label{eq:vb}
V_{\mathbf{B}}(\mathbf{r})
&
= V_1 + \nonumber\\
&2V_0\sum_{\alpha=1}^3
\cos\left[\mathbf{G}^{\mathrm{M}}_\alpha\!\cdot\!\mathbf{r}
    + \bar{\mathbf{b}}_\alpha\!\cdot\!(\mathbf{u}^{(2)}(\mathbf{r})\!-\!\mathbf{u}^{(1)}(\mathbf{r})) + \varphi_0\right]  
        ,
\end{align}
Here, we defined
$\mathbf{G}_{\alpha}^{\mathrm{M}}=\mathbf{b}_\alpha-\mathbf{b}_\alpha'$ and
$\bar{\mathbf{b}}_{\alpha}=\mathbf{b}_\alpha+\mathbf{b}_\alpha'$,
where $\mathbf{b}_\alpha$ and $\mathbf{b}_\alpha'$ ($\alpha=1,2,3$) denote the three primitive reciprocal lattice vectors of graphene and hBN, respectively, forming a trigonal set with $120^\circ$ rotations.
The parameters are given by
$V_0=0.202~\mathrm{eV/nm^2}$, $V_{\mathrm{1}}=-0.700~\mathrm{eV/nm^2}$,
and $\varphi_0=0.956$~\cite{krisna2025low, san2014spontaneous},
which yield the relation 
$V_{\mathrm{AB'}}<V_{\mathrm{BA'}} < V_{AA'}$.
For an infinite 2D graphene/hBN superlattice, the relaxed displacement field $\mathbf{u}^{(l)}(\mathbf{r})$ is obtained by minimizing Eq.~\eqref{eq:eplusb} under two-dimensional periodic boundary conditions, resulting in a hexagonal array of the energetically most stable AB$'$ domains~\cite{san2014spontaneous,jung2015origin,krisna2023moire,krisna2025low}.



The continuum model cannot be directly applied to a GNR on hBN, due to the ambiguity of implementing the edge boundary condition for
$\mathbf{u}^{(l)}(\mathbf{r})$.
To describe the relaxation of the finite-width system, we employ a discrete square-grid model, illustrated in Fig.~\ref{grid}(b), which reduces to Eq.~\eqref{eq:eplusb} in the continuum limit.
In this model, both the graphene and hBN layers are represented by the  parallel square grids with the same spacing $l_0$ and the same orientation regardless of the twist angle.
The vertices of the grid represent mass points,
and springs are assigned along orthogonal and diagonal directions.
Note that the vertices do not coincide with the actual atomic positions.

The grid spacing $l_0$ is chosen to be sufficiently small compared to the characteristic scale of lattice relaxation.
Specifically, we divide the one-dimensional superlattice period $\Lambda_{\mathrm{M}}$ by an integer $M_x$ such that $l_0 \sim 0.5$ nm, i.e.,
$M_x = [\Lambda_{\mathrm{M}}/0.5\mathrm{nm}]$
where $[x]$ denotes the nearest integer to $x$.
A periodic boundary condition is imposed along the $x$ direction with period $M_x$.
Along the $y$ direction, the GNR width is discretized into $M_y$ grids with open boundary conditions, 
where we set $M_y =$ 4, 9, 17, 26 to approximate the case of $N =$ 10, 20, 40, 60, respectively.
The hBN layer is modeled as a wider strip of width $M_y+10$, providing a margin of $5$ on each side.
We label the 2D grid points before relaxation by $\mathbf{r}_i$, where graphene and hBN share the same set of $\mathbf{r}_i$ within the overlap region.


For the defined grids,
we consider the in-plane displacement $\mathbf{u}^{(l)}(\mathbf{r}_i)$ for the mass point
of the original position $\mathbf{r}_i$ on layer $l$,
as illustrated in Fig.~\ref{grid}(b).
The total energy in the presence of displacement is written as a function of $\{\mathbf{u}^{(l)}(\mathbf{r}_i)\}$,
which is the sum of elastic and binding contributions:
\begin{equation}
U^{\mathrm{(eff)}}=U^{\mathrm{(eff)}}_{\mathrm{E}}+U^{\mathrm{(eff)}}_{\mathrm{B}}.
\end{equation}
The elastic part is given by
\begin{align}\label{eq:ueeff}
U^{\mathrm{(eff)}}_{\mathrm{E}} =
\sum_{l=1,2}\Bigg[\frac{k^{(l)}}{2}\sum_{\langle i,j\rangle} \delta d_{i,j}^{(l)2}
+\frac{k^{(l)}_{\mathrm{d}}}{2}\sum_{\langle\!\langle i,j\rangle\!\rangle} \delta d_{i,j}^{(l)2}
\nn\\+\frac{\kappa^{(l)}}{2l_0^2} \sum_i (S_{i}^{(l)}-l_0^2)^2 \Bigg],
\end{align}
where
$\langle i,j \rangle$ denotes atomic pairs connected by vertical and horizontal bonds, while $\langle\!\langle i,j \rangle\!\rangle$ refers to pairs connected by diagonal bonds.
The $\delta d_{i,j}^{(l)}$ is the change in bond distance given by
\begin{equation}\label{eq:deld}
    \delta d_{i,j}^{(l)} = |\mathbf{r}_i+\mathbf{u}^{(l)}(\mathbf{r}_i)-\mathbf{r}_j-\mathbf{u}^{(l)}(\mathbf{r}_j)|-|\mathbf{r}_i-\mathbf{r}_j|,
\end{equation}
and $S^{(l)}_i$ denotes the area of the deformed square plaquette with grid point $i$ at its lower-left corner [See Fig.~\ref{grid}(b)].
For the correspondence with the coninuum model Eq.~\eqref{eq:vecont}, we define the force constants as
\begin{align}\label{eq:kvslame}
k^{(l)} = 2\mu^{(l)}, \quad
k_d^{(l)} = \mu^{(l)}, \quad
\kappa^{(l)} = \lambda^{(l)}-\mu^{(l)}.
\end{align}
We can show that the elastic energy $U_{\rm E}^{\rm (eff)}$ asymptotically agrees with Eq.~\eqref{eq:vecont}
in the continuum limit [see Appendix \ref{app_grid}].

Corresponding to Eq.~\eqref{eq:vb}, the binding energy of the sping-mass model is written as
\begin{align}\label{eq:vbeff}
U^{\mathrm{(eff)}}_{\mathrm{B}} = {\sum_{i\in\mathrm{overlap}}} V_{\mathrm{B}}(\mathbf{r}_i)l_0^2
\end{align}
where 
$V_{\mathrm{B}}(\mathbf{r})$ is given by Eq.~\eqref{eq:vb},
and the summation of $\mathbf{r}_i$ runs only over the grid points in the overlapped area of GNR and hBN.
Note that the dependence of the moir\'e pattern on the twist angle is encoded in $\mathbf{G}^{\rm M}_\alpha$ within the function $V_{\rm B}(\mathbf{r})$.

We numerically compute the relaxed displacement field $\mathbf{u}^{(l)}(\mathbf{r}_i)$ 
on the square grid using the steepest descent method. 
At each iteration, the displacement is updated as 
\begin{equation}
u_{\alpha,\mathrm{new}}^{(l)}(\mathbf{r}_i)
=u_{\alpha,\mathrm{old}}^{(l)}(\mathbf{r}_i)
-\eta \frac{\partial U^{(\mathrm{eff})}}{\partial u_{\alpha}^{(l)}(\mathbf{r}_i)}\Bigg|_{\mathbf{u}^{(l)}=\mathbf{u}_{\mathrm{old}}^{(l)}}
\end{equation}
with a sufficiently small step size $\eta$.
The iteration continues until $U^{\mathrm{(eff)}}$ converges. 
To ensure the global minimum, we initialize the displacement 
field with several trial configurations and 
select the solution with the lowest energy. 
In the numerical relaxation, we fix the original periodicity $\Lambda_{\rm M}$ along the $x$ direction, which effectively simulates the situation where both ends of the ribbon are pinned to the substrate.

\begin{figure*}
        \begin{center}
        \leavevmode\includegraphics[width=1. \hsize]
        {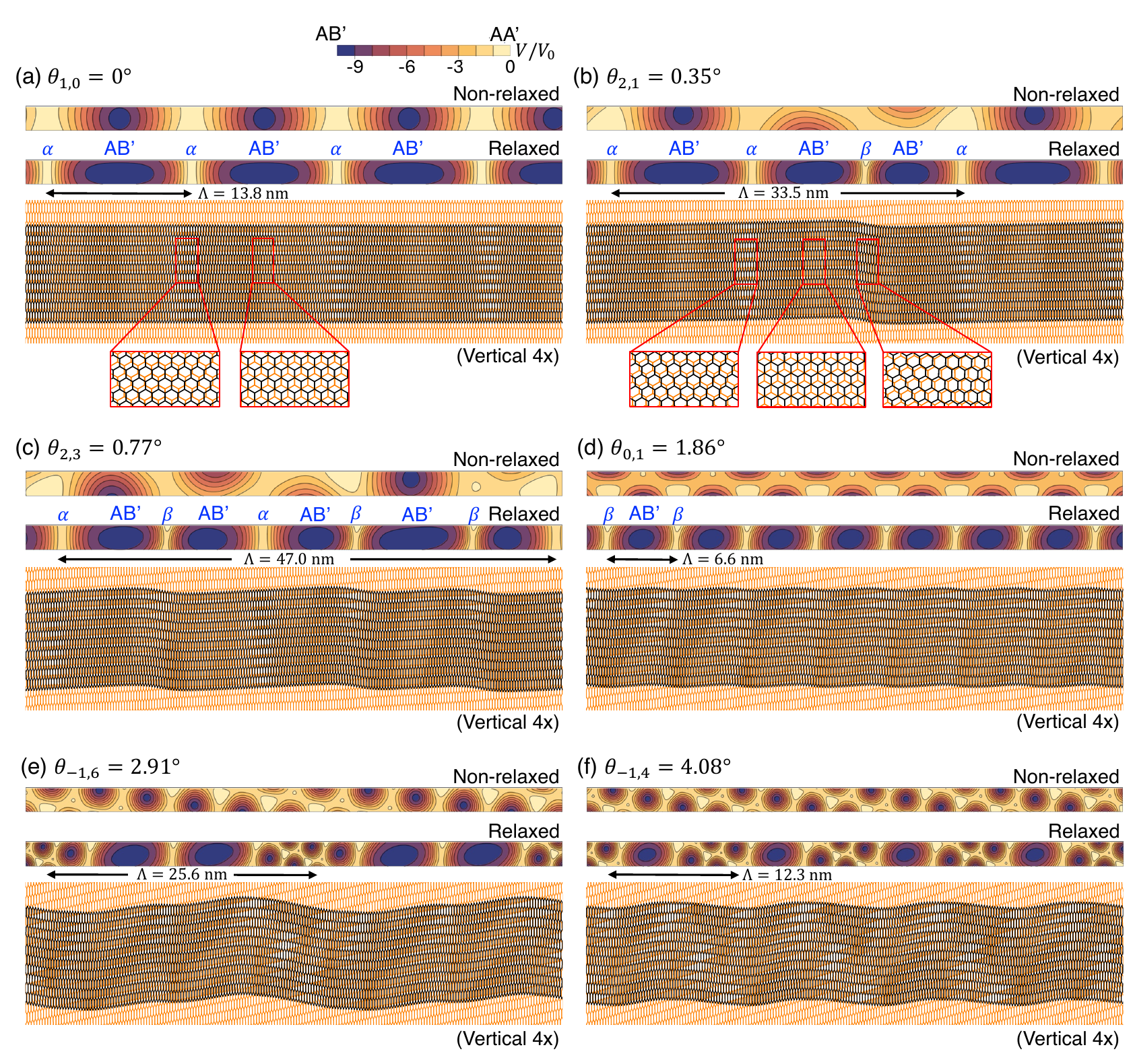}
    \caption{
        Optimized structures of zigzag GNRs with width $N=10$ on hBN at twist angles (a) $\theta_{1,0}=0^{\circ}$, (b) $\theta_{2,1}=0.35^{\circ}$, (c) $\theta_{2,3}=0.77^{\circ}$, (d) $\theta_{0,1}=1.86^{\circ}$, (e) $\theta_{-1,6}=2.91^{\circ}$, and (f) $\theta_{-1,4}=4.08^{\circ}$. In each panel, the top and middle plots show the interlayer binding energies before and after relaxation, respectively, while the bottom panel shows the corresponding relaxed lattice structure with the vertical axis magnified by a factor of four. The symbols $\alpha$ and $\beta$ denote the two types of domain walls (see text).
        }
        \label{optimized_energy_structure}
        \end{center}
    \end{figure*}

\section{Optimized structure}
\label{sec:OLS}


Figure~\ref{optimized_energy_structure} summarizes the optimized structures of GNRs with width $N=10$ on hBN at twist angles $\theta=0^{\circ}, 0.35^{\circ}, 0.77^{\circ}, 1.86^{\circ}, 2.91^{\circ}$, and $4.08^{\circ}$.
In each panel, the upper two figures 
present contour maps of the interlayer binding energy 
of non-relaxed and relaxed structures, respectively,
mapped on the original non-relaxed space, $\bm{r}_i$.
The lower panel shows the corresponding relaxed lattice structure, with the $y$ direction magnified by a factor of four for clarity. The atomic positions of the graphene and hBN honeycomb lattices are obtained by interpolating the displacement fields defined on the effective square grid.
The horizontal black arrow represents the 1D superlattice period $\Lambda$.

At $\theta=0^{\circ}$ [Fig.~\ref{optimized_energy_structure}(a)], upon relaxation, the system expands the most stable AB$^\prime$-stacking regions in order to achieve the minimum interlayer binding energy, forming a one-dimensional domain structure. 
Between neighboring AB$^\prime$ domains, a domain wall emerges due to the lattice constant mismatch along the $x$ direction, where the graphene and hBN lattices are relatively shifted in the horizontal direction. In the following, we refer to the domain wall of this structure as the $\alpha$ type.
The width of the domain wall is on the order of a few nanometers, consistent with that in 2D graphene/hBN superlattices, and is determined by the balance between interlayer binding and elastic energies \cite{nam2017lattice,carr2018relaxation,krisna2023moire}.

At $\theta = 0.35^{\circ}$ [Fig.~\ref{optimized_energy_structure}(b)], three domain walls appear within a single superlattice period. Two of these are of the $\alpha$ type described above. In both the $\alpha$ domain walls and AB$'$ domains, the horizontal rows of hexagons in graphene and hBN remain aligned in parallel, as in the $\theta = 0^\circ$ case. In contrast, the remaining domain wall has a distinct configuration, referred to as the $\beta$ type, in which the horizontal row of graphene is vertically shifted to an adjacent lane of hBN, to satisfy the periodic boundary condition under the rotated configuration.
As the twist angle increases, additional $\beta$ domains emerge to accommodate the larger relative orientation between graphene and hBN, resulting in a wavy ribbon profile.
Notably, the numbers of $\alpha$ and $\beta$ domain walls per $\Lambda$ are given by $m$ and $n$, respectively.
For example, at $\theta = 0.77^{\circ}$ with $(m,n)=(2,3)$, we observe two $\alpha$ domains and three $\beta$ domains per period [Fig.~\ref{optimized_energy_structure}(c)], whereas at $\theta = 1.86^{\circ}$ with $(m,n)=(0,1)$, there is a single $\beta$ domain per period [Fig.~\ref{optimized_energy_structure}(d)].
This relation between $(m,n)$ and the number of domain walls is discussed in more detail in Appendix~\ref{sec:moire_mapping}.
We note that these characteristic moir\'e pattern relaxations are specific to the 1D GNR system and do not correspond to any portion of the relaxed structure of a 2D graphene/hBN superlattice.

At higher twist angles, $\theta = 2.91^{\circ}$ and $4.08^{\circ}$ [Fig.~\ref{optimized_energy_structure}(e) and (f)], the orientation mismatch becomes too large for the system to accommodate all domains and domain walls uniformly. In this regime, relaxation selectively enlarges certain AB$^\prime$ regions while compressing the remaining patterns over short distances. At even higher angles, this selective domain formation disappears entirely, and the structure approaches the non-relaxed, nearly uniform configuration.

While our analysis focuses on the commensurate cases labeled by $(m,n)$, a general GNR/hBN system is incommensurate and quasi-periodic. In such quasi-periodic configurations, we expect the emergence of a similar 1D moir\'e structure with AB$'$ domains separated by $\alpha$ and $\beta$ domain walls, although the sequence of domains and domain walls is no longer strictly periodic.

\begin{figure*}
        \begin{center}
        \leavevmode\includegraphics[width=1. \hsize]
        {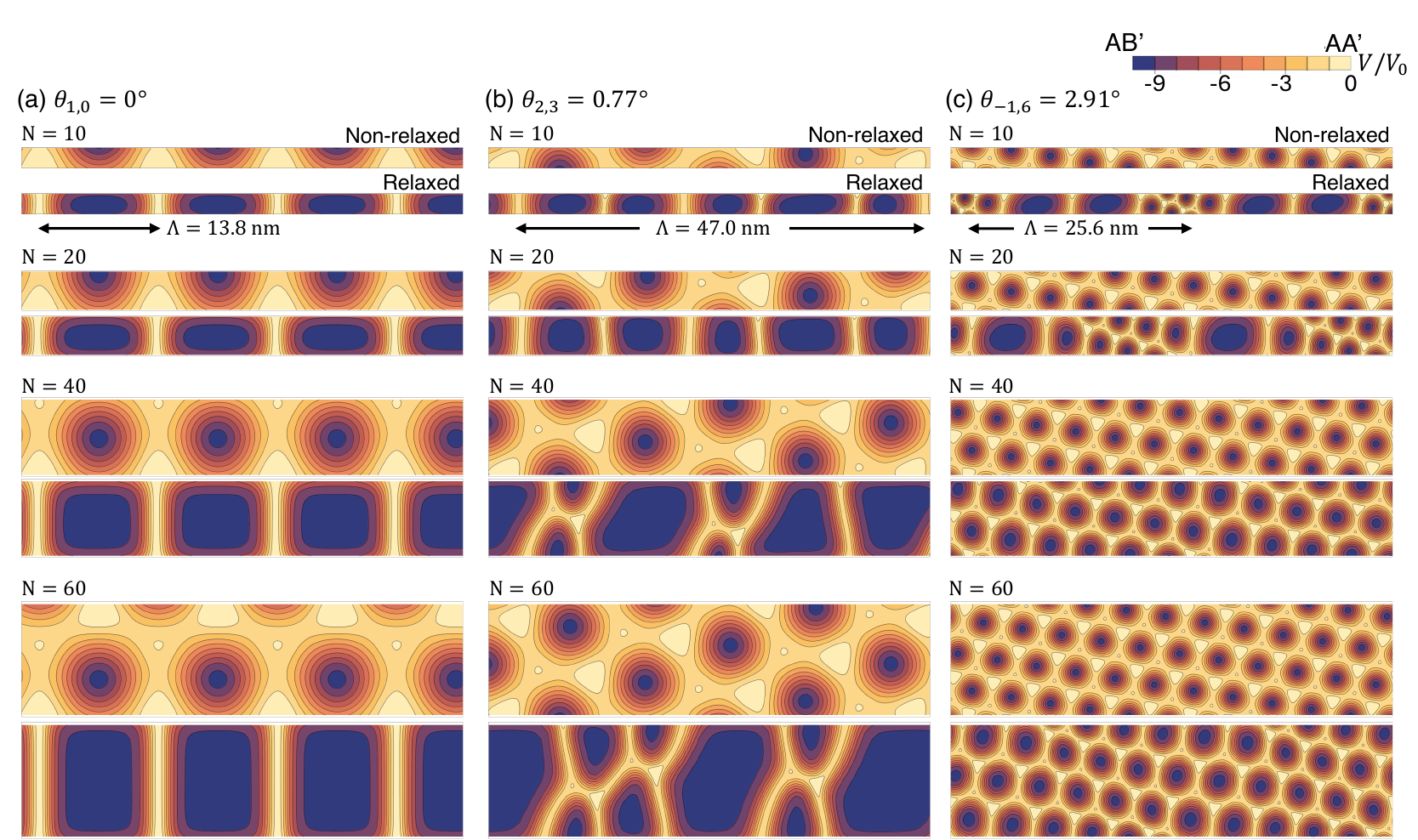}
    \caption{
        Interlayer binding energy of GNR on hBN, similar to Fig.~\ref{optimized_energy_structure}, at twist angles (a) $\theta_{1,0}=0^{\circ}$, (b) $\theta_{2,3}=0.77^{\circ}$, and (c) $\theta_{-1,6}=2.91^{\circ}$, for ribbon widths $N=10, 20, 40,$ and $60$. In each panel, the upper and lower plots correspond to the non-relaxed and relaxed structures, respectively.
        }
        \label{width_dep}
        \end{center}
    \end{figure*}


Figure~\ref{width_dep} shows contour maps of the optimized interlayer binding energy for GNRs of differnet widths $N = 10, 20, 40,$ and $60$, at selected twist angles $\theta$. In each panel, the upper and lower figures correspond to the non-relaxed and relaxed configurations, respectively.
At $\theta = 0^\circ$, widening the ribbon simply enlarges the AB$'$ stacking regions vertically, thereby preserving the characteristic 1D domain structure [Fig.~\ref{width_dep}(a)].

At $\theta = 0.77^\circ$ [Fig.~\ref{width_dep}(b)], narrow ribbons ($N=10$ and $20$) still maintain a one-dimensional domain arrangement similar to that at $\theta=0^\circ$. However, for wider ribbons ($N=40$ and $60$), the simple 1D alignment breaks down, and the relaxation produces H-shaped domain-wall patterns. This indicates a crossover from a purely 1D configuration to a mixed state that incorporates two-dimensional features.
At larger twist angles, such as $\theta = 2.91^\circ$ [Fig.~\ref{width_dep}(c)], the relaxed structures gradually evolve into two-dimensional domain patterns, similar to those found in 2D graphene/hBN system.

In GNRs wider than the 2D moir\'e period $L_{\rm M}$, the relaxation process can yield multiple metastable structures depending on the initial configuration. 
For example, in the case of $N=60$ at $\theta=0^\circ$, 
the 1D structure shown in Fig.~\ref{width_dep}(a), which has the lowest energy among systems with fixed $\Lambda_{\rm M}$, is obtained when the initial state is prepared by expanding the GNR width to match the hBN lattice constant. 
In contrast, when we start from the non-relaxed configuration (a simple overlap of the intrinsic GNR on hBN), we obtain a metastable state with a 2D hexagonal moir\'e motif. We expect the latter condition to be more representative of experimental situations where a GNR is simply placed on the hBN substrate, while the former ground state may be realized through an appropriate annealing process. Finally, if we remove the boundary condition along the $x$ direction, the true lowest-energy configuration is a fully commensurate structure with uniform AB$'$ stacking.

\section{Electronic structure}
\label{sec:ES}

\subsection{Tight-binding model}
\label{subsec:TB model}

The electronic structure of the relaxed GNR/hBN superlattice is analyzed within a tight-binding framework.
Let $\mathbf{R}^{0}_i$ denote the atomic position of site $i$ on the unstrained honeycomb lattice of
the GNR and hBN layers. 
The relaxed atomic positions are given by
\begin{equation}
\mathbf{R}_i = \mathbf{R}_{i}^{0} + \mathbf{u}(\mathbf{R}_i^{0}),
\end{equation}
where the displacement field $\mathbf{u}(\mathbf{R}_i^0)$ is obtained
by interpolating $\mathbf{u}^{(1)}(\mathbf{r})$ for graphene and $\mathbf{u}^{(2)}(\mathbf{r})$ for hBN,
as discussed in Sec.~\ref{sec:OLS}.
We neglect the out-of-plane displacement and fix the interlayer spacing at a constant value of $d \approx 0.334~\mathrm{nm}$.

In the tight-binding model, we include the $p_z$ orbitals of carbon atoms in graphene and boron/nitrogen atoms in hBN. 
The Hamiltonian is written as a summation of 
the intralayer matrices $H_{\rm G}, H_{\rm hBN}$ and the interlayer matrix $T$,
as
\cite{moon2014electronic}
\begin{align}
\label{eq5}
    &H = H_{\rm G} + H_{\rm hBN} + T,
    \end{align}
with
\begin{align}
   & H_{\rm G} = -\sum_{i\neq j} t_{\rm G}(\mathbf{R}_i-\mathbf{R}_j) c_i^\dag c_j, \quad  H_{\rm hBN} = \sum_{i} V_i d_i^\dag d_i,
     \nonumber\\
    &  T_{\rm G-hBN} =  -\sum_{i,j} t_{\rm G-hBN}(\mathbf{R}_i-\mathbf{R}_j) c_i^\dag d_j+ {\rm h.c.},
\end{align}
where $c_i$ and $d_i$ denote the annihilation operator 
on site $i$ of graphene and hBN layer, respectively.
For intralayer hopping on graphene and interlayer hopping, we assume the same function 
$t(\mathbf{R}) \equiv t_{\rm G}(\mathbf{R}) = t_{\rm G-hBN}(\mathbf{R})$ define by
\begin{gather}\label{eq6}
        -t(\mathbf{R})=V_{p\!p\!\pi}\left[1-\left(\frac{\mathbf{R}\cdot\mathbf{e}_z}{R}\right)^2\right]
        +V_{p\!p\!\sigma}\left(\frac{\mathbf{R}\cdot\mathbf{e}_z}{R}\right)^2,\\
        V_{p\!p\!\pi}=V_{p\!p\!\pi}^0e^{-(R-a/\sqrt{3})/r_0}, \
        V_{p\!p\!\sigma}=V_{p\!p\!\sigma}^0 e^{-(R-d)/r_0},\notag
\end{gather}
where $\mathbf{e}_z$ is the unit vector along $z$ axis, and we adopt $V_{p\!p\!\pi}^0 \approx -2.7~\mathrm{eV}$, $V_{p\!p\!\sigma}^0 \approx 0.48~\mathrm{eV}$, and $r_0=0.184a$ \cite{moon2014electronic}.
The $V_i$ is the on-site potential on hBN layer, which  is set as \cite{slawinska2010energy}
\begin{equation}
\label{eq_VB_VN}
    V_{\rm B} = 3.34 \, {\rm eV} \, \mbox{(boron)}, \quad 
    V_{\rm N} = -1.40 \, {\rm eV} \, \mbox{(nitrogen)}
\end{equation}
We neglect the intralayer hoppings within hBN layer, which has a little effect on the low-energy band of graphene \cite{moon2014electronic}.

The electronic states are obtained by solving the Schrödinger equation in the 1D unit cell of length $\Lambda$ with Bloch momentum $k$.
The local density of states at site $\mathbf{R}_i$
and energy $E$ is calculated as
\begin{equation}
 \rho(\mathbf{R}_j,E) = \sum_{n,k} |\bm{\psi}_{nk}(\mathbf{R}_j)|^2\delta(E-E_{nk})
\end{equation}
where $\bm{\psi}_{nk}(\mathbf{R}_j)$ is the wave amplitude on site $j$ for the $n$th eigenstate at wavenumber $k$
and $E_{nk}$ is the eigenenergy.
The delta function is approximated by a Lorentzian, $\delta(E) \approx \eta / (E^2 + \eta^2)$, with a small broadening factor $\eta = 0.01~\mathrm{meV}$.

For interpreting the LDOS results in the following, it is useful to derive the effective potential for the zigzag edge states, arising from the second-order process of the interlayer coupling.
The effective Hamiltonian of graphene electrons including the effect of hBN can be derived by integrating out the hBN states as \cite{moon2014electronic}
\begin{align}
H^{\rm (eff)}_{G} = H_{\rm G} + \Delta H,
\quad
\Delta H = T^\dagger(-H_{\rm hBN})^{-1}T.
\end{align}
The space of the zero-energy zigzag edge states on a given edge (top or bottom) is approximately spanned by wave packets whose amplitudes reside only on the outermost edge sites with alternating signs.
By projecting $\Delta H$ onto this edge-state subspace, we obtain the effective on-site potential as
\begin{equation}
U_i=
\langle i |\Delta H|i\rangle
- {\rm Re}\, \bigl[
\langle i+1 |\Delta H|i\rangle
+\langle i-1 |\Delta H|i\rangle
\bigr],
\label{eq_U_i}
\end{equation}
where $i$ labels the outermost sites.
The negative sign in the second term originates from the opposite signs of the edge-state wavefunction at sites $i$ and $i\pm1$.
We neglect long-distance hopping $\langle j |\Delta H| i\rangle$ with $|j-i|\ge 2$, as they are negligibly small.
As will be shown in the following section, the resulting $U_i$ varies smoothly along the ribbon axis, leading to a pronounced modulation of the low-energy spectrum and lifting the degeneracy of the zigzag edge modes.

\begin{figure*}
        \begin{center}
        \leavevmode\includegraphics[width=1 \hsize]
        {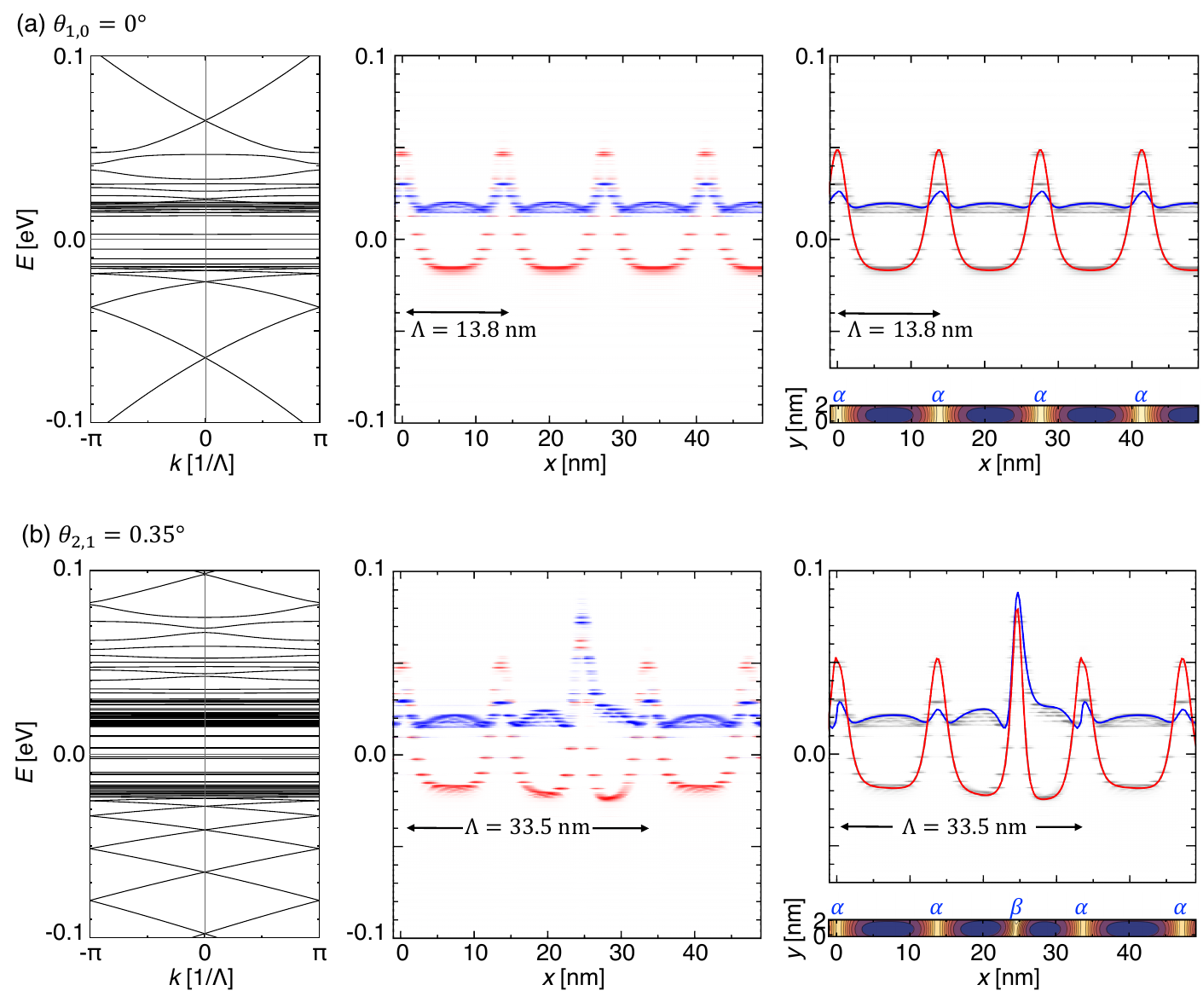}
    \caption{
        Electronic structures of GNR/hBN at twist angles (a) $\theta_{1,0}=0^{\circ}$ and (b) $\theta_{2,1}=0.35^{\circ}$. From left to right: band structure, LDOS (red for upper edge, blue for lower edge), and LDOS overlaid with the effective on-site potential (shown in gray). The lower-right panel shows the contour map of the interlayer binding energy, corresponding to Fig.~\ref{optimized_energy_structure}.
        }
        \label{electronic_states_0deg_0.35deg}
        \end{center}
    \end{figure*}

\subsection{Electronic structure}

Figure~\ref{electronic_states_0deg_0.35deg} summarizes the electronic structures of the relaxed $N=10$ GNR for (a) $\theta_{1,0}=0^\circ$ and (b) $\theta_{2,1}=0.35^\circ$.
In each panel, the left figure displays the band structure, 
and the middle panel presents the LDOS as a density map in $x$–$E$ space, where red and blue correspond to the upper and lower edges, respectively.
The right figure presents the same LDOS in gray, overlaid with the effective edge-site potential $U_i$ [Eq.~\eqref{eq_U_i}] for the upper (red) and lower (blue) edges.
The lower-right panel illustrates the contour map of the interlayer binding energy, corresponding to Fig.~\ref{optimized_energy_structure}.

In both cases, all the energy bands within the plotted region originate from zigzag edge states, which are strongly modulated by the hBN potential. The corresponding LDOS closely follows the effective potential $U_i$. We find that $U_i$ is nearly constant within the AB$'$ domains but exhibits pronounced peaks at the domain boundaries. The values of $U_i$ in the AB$'$ domain differ between the upper and lower edges by about 40 meV, because the upper edge sites are nearly aligned on top of boron atoms, whereas the lower edge sites lie above the centers of the hBN hexagons. Since boron atoms have a positive on-site energy $V_B$ [Eq.~\eqref{eq_VB_VN}], this alignment lowers the effective potential of the graphene sites through a level-repulsion effect.

Consequently, the energy subbands are densely distributed within the energy range corresponding to the domain potential plateaus, whereas the potential peaks at the boundaries give rise to sparsely distributed states in the energy windows between the red and blue plateaus.
Each of these sparsely distributed states possesses a nearly constant spatial width of about 1 nm. This corresponds to the minimum width of an edge-state wave packet, $w \sim 2\pi/\Delta k$, where $\Delta k = 2\pi/(3a)$ represents the momentum-space range of the edge-state flat band \cite{fujita1996peculiar, nakada1996edge, son2006energy, son2006half}.
We also note that the potential peaks at the domain walls are higher in $\beta$ than in $\alpha$, which is reflected in the LDOS distribution.

\begin{figure*}
        \begin{center}
        \leavevmode\includegraphics[width=1. \hsize]
        {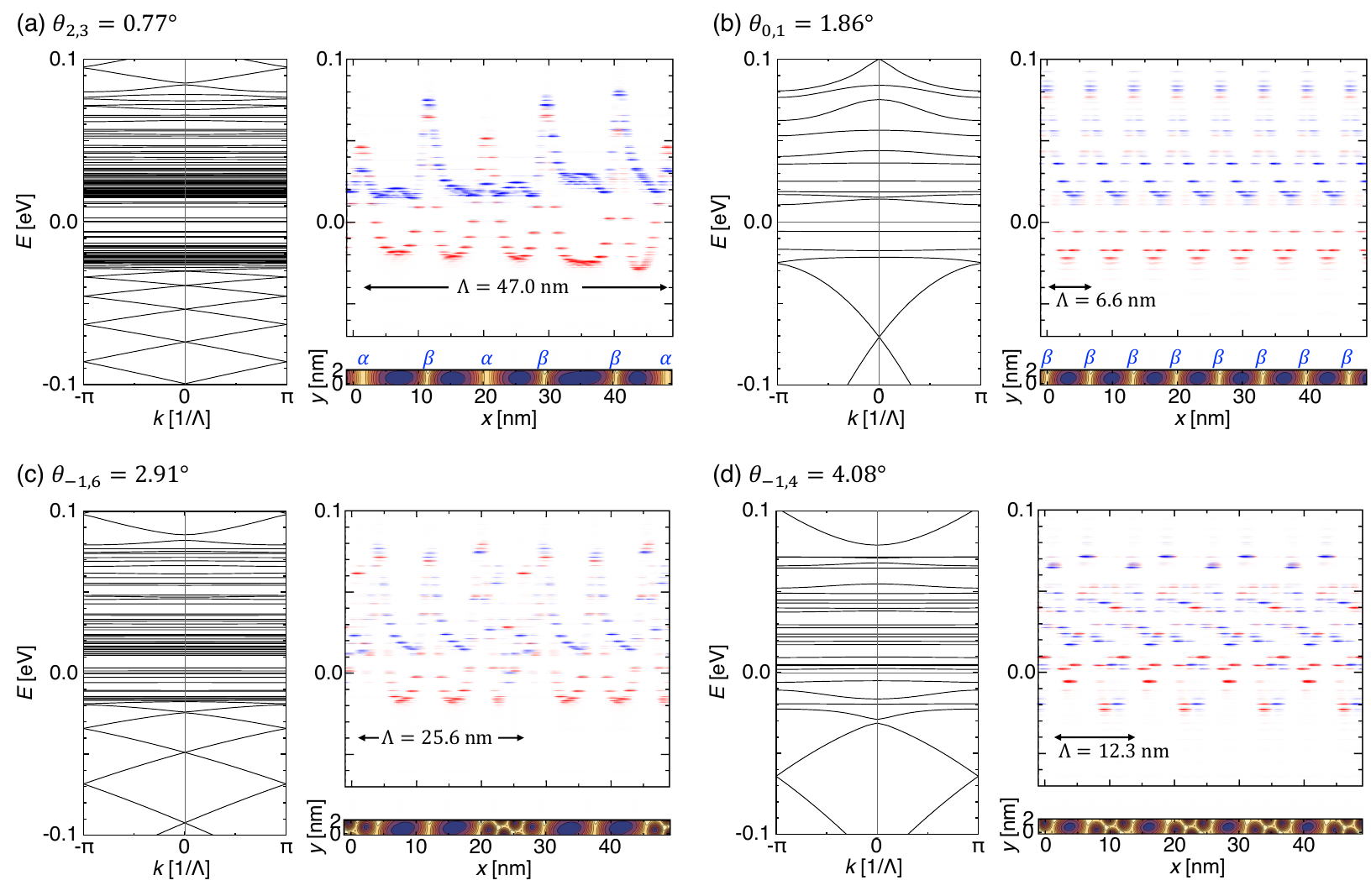}
    \caption{
        Electronic structures of GNR/hBN at twist angles (a) $\theta_{2,3}=0.77^{\circ}$, (b) $\theta_{0,1}=1.86^{\circ}$, (c) $\theta_{-1,6}=2.91^{\circ}$, and (d) $\theta_{-1,4}=4.08^{\circ}$. For each case, the left panel shows the band structure, and the right panel shows the LDOS (red for upper edge, blue for lower edge). The lower-right panel displays the corresponding relaxed interlayer binding energy in Fig.~\ref{optimized_energy_structure}.
        }
        \label{electronic_states}
        \end{center}
\end{figure*}

Lastly, we show the band structure and LDOS of the $N=10$ GNR at other twist angles in Fig.~\ref{electronic_states}, where the same trends are observed. At higher twist angles [Fig.~\ref{electronic_states}(c) and (d)], the separation of domain and domain-wall states is prominent only in the well-formed domains, while the states are mixed in the remaining short-moir\'e-period regions.

The present results suggest that the GNR/hBN heterostructure offers a unique platform for realizing one-dimensional arrays of quantum-confined electronic states. When the Fermi energy lies within the energy window between the states localized in different stacking domains, electronic states emerge at the domain walls, forming a periodic chain of one-dimensional quantum dots with a spacing on the order of the moir\'e period. Owing to the short confinement length set by the atomic-scale edge and domain-wall potentials, strong Coulomb interactions and pronounced single-electron charging effects are expected. A small shift of the Fermi level by gate tuning can relocate the carrier localization from the domain walls to the domain centers, suggesting a possibility of electrostatically switchable localization. At higher doping levels, the system can abruptly transition to a conductive regime through the dispersive subbands. Compared with armchair nanoribbons, the zigzag edge provides a much sharper contrast between localized and conducting states, enabling a highly tunable and structurally well-defined one-dimensional quantum-dot array formed simply by stacking the two lattices.


\section{Conclusion}
\label{sec:conclusion}

We have investigated the structural relaxation and electronic properties of zigzag graphene nanoribbons placed on hexagonal boron nitride substrates, forming a one-dimensional moir\'e system.
By employing an effective grid model derived from continuum elasticity theory, we clarified how the atomic structure relaxes into a periodic sequence of commensurate AB$'$ domains separated by domain walls.
At finite twist angles, the nanoribbon adopts a wavy configuration in which the ribbon primarily follows the zigzag orientation of the hBN lattice, while occasionally undergoing lateral shifts to adjacent atomic lanes. This gives rise to a characteristic one-dimensional domain structure with two distinct types of domain boundaries
referred to as $\alpha$ and $\beta$, corresponding to relative atomic displacements along the ribbon and perpendicular to it, respectively, allowing the system to accommodate the rotational misalignment.
Tight-binding calculations revealed that the moir\'e-induced potential strongly modulates the zero-energy edge states, producing domain-localized subbands and sharply confined domain-wall states.
These findings highlight the critical role of structural relaxation in defining the electronic landscape of 1D moir\'e systems and suggest new possibilities for edge-state engineering and the design of moir\'e-based nanodevices.

\section*{Acknowledgments}

The authors are grateful to R. Nakayama for stimulating discussions.
This work was supported in part by JSPS KAKENHI Grants No. JP25K00938, No. JP21H05236, No. JP21H05232, No. JP24K06921 and by JST CREST Grant No. JPMJCR20T3, Japan. 
N.N. also acknowledges
the support from the JSPS Overseas Research Fellowship.


\appendix
\section{Derivation of the effective grid model}
\label{app_grid}

In this Appendix, we show that the energy expression of the effective discrete model in Eq.~\eqref{eq:ueeff} reduces to that of the continuum model, Eq.~\eqref{eq:vecont}, in the limit of vanishing grid spacing $l_0 \to 0$.
The effective discrete model is composed of two layers of a square grid
($l=1,2$) as illustrated in Fig.~\ref{grid},
where the vertexes are connected by springs in the horizontal, vertical, and diagonal directions with elastic constants $k^{(l)}$, $k^{(l)}$, and $k_{\rm d}^{(l)}$, respectively.
For a small and smoothly varying displacement field $\mathbf{u}^{(l)}(\mathbf{r})$, the change in the bond length [Eq.~\eqref{eq:deld}] can be approximated to first order as
\begin{align}
\label{eq:deld2}
\delta d_{i,j}^{(l)} &= \big|\mathbf{r}_{ij}+\mathbf{u}^{(l)}_{ij}\big|- \big|\mathbf{r}_{ij}\big| \,\,\approx \frac{\mathbf{r}_{ij}}{|\mathbf{r}_{ij}|}\cdot \mathbf{u}^{(l)}_{ij} \nonumber\\ &\approx \frac{\mathbf{r}_{ij}}{|\mathbf{r}_{ij}|}\cdot (\mathbf{r}_{ij}\cdot \nabla) \mathbf{u}^{(l)}(\mathbf{r}_i), 
\end{align} 
where $\mathbf{r}_{ij}=\mathbf{r}_j-\mathbf{r}_i$, $\mathbf{u}^{(l)}_{ij}=\mathbf{u}^{(l)}(\mathbf{r}_j)-\mathbf{u}^{(l)}(\mathbf{r}_i)$.
In the last line, we used the approximation
$\mathbf{u}^{(l)}_{ij} \approx (\mathbf{r}_{ij}\cdot\nabla)\mathbf{u}^{(l)}(\mathbf{r}_i)$,
which is valid for a smoothly varying displacement field.

By applying Eq.~\eqref{eq:deld2} to the bonds along the three directions, we obtain
\begin{equation}
\delta d_{ij}^{(l)}\simeq
\left\{\begin{array}{cl}
 l_0 u_{xx}^{(l)}  & \hbox{for }\Delta{\mathbf{r}}_{ij} = l_0\mathbf{e}_x, \\
 l_0 u_{yy}^{(l)} & \hbox{for }\Delta{\mathbf{r}}_{ij} = l_0 \mathbf{e}_y,\\
\dfrac{l_0}{\sqrt{2}} \Big(u_{xx}^{(l)} + u_{yy}^{(l)} \pm 2u_{xy}^{(l)}\Big)
& \hbox{for } \Delta{\mathbf{r}}_{ij} =  l_0(\pm\mathbf{e}_x \blue{+} \mathbf{e}_y).
\end{array}
\right.
\end{equation}
Similarly, the change in the area of a square plaquette, $S_{i}^{(l)}$, is written to leading order of $l_0$ as
\begin{equation}
S_i^{(l)} - l_0^2 \simeq l_0^2 \left[u_{xx}^{(l)}(\mathbf{r}_i) + u_{yy}^{(l)}(\mathbf{r}_i)\right].
\end{equation}
By using these relations, 
the effective elastic energy in Eq.~\eqref{eq:ueeff} becomes
\begin{align}
U_{\mathrm{E}}^{(\mathrm{eff})}
\simeq
&\sum_{l=1,2}\int d^2\mathbf{r}\,
\Bigg\{
\frac{k^{(l)}}{2}\left[(u_{xx}^{(l)})^2+(u_{yy}^{(l)})^2\right]\nonumber\\
&+\frac{k_d^{(l)}}{2}\!
\left[
(u_{xx}^{(l)}+u_{yy}^{(l)})^2+(2u_{xy}^{(l)})^2\right] \nonumber\\
&+\frac{\kappa^{(l)}}{2}\left(u_{xx}^{(l)}+u_{yy}^{(l)}\right)^2
\Bigg\},
\end{align}
where we replaced the summation $\sum_{i} l_0^2$ to the integral $\int d^2\mathbf{r}$. 
By comparing this expression with the continuum elastic energy in Eq.~\eqref{eq:vecont}, 
we identify the corresondence between the force constants and the Lam\'e parameters, 
as summarized in Eq.~\eqref{eq:kvslame}

\section{
Moir\'e map interpretation of the superlattice domain structure
}

\label{sec:moire_mapping}

In the main text, we found that the numbers of $\alpha$ and $\beta$ domain walls within a single superlattice period $\Lambda$ correspond to the indices $m$ and $n$, respectively.
This relation can be understood through the moir\'e mapping illustrated below.

Figure~\ref{spider_map}(a) shows the moir\'e pattern of a non-relaxed graphene/hBN structure in the left panel, where the black grid represents the effective square lattice for a GNR with $N=10$.
The right panel presents the relaxed structure mapped onto the same, non-relaxed moir\'e pattern.
This mapping is obtained by assigning each grid point of the relaxed ribbon to the position in the non-relaxed moir\'e pattern that has the same local interlayer registry.
The lower-right panel displays the corresponding moir\'e pattern plotted on the undeformed grid [as in Fig.~\ref{optimized_energy_structure}(c)].

We observe that a large portion of the ribbon area is concentrated near the AB$'$ stacking regions in the map, indicating the formation of AB$'$ domains.
Each domain wall corresponds to a connection between neighboring AB$'$ spots.
Here, the $\alpha$ and $\beta$ domain walls correspond to displacements along the $\mathbf{L}^\mathrm{M}_1$ and $\mathbf{L}^\mathrm{M}_2$ directions, respectively.
This correspondence arises because a translation in the moir\'e contour map by $\mathbf{L}^\mathrm{M}_i$ represents a change in the interlayer sliding (of hBN relative to graphene) by $\mathbf{a}_i$.
In the $\alpha$ domain, the local interlayer sliding occurs along $\mathbf{a}_1$, corresponding to a displacement by $\mathbf{L}^\mathrm{M}_1$ in the moir\'e map.
Similarly, the $\beta$ domain corresponds to a displacement along $\mathbf{L}^{\mathrm{M}}_{2}$.
Consequently, a single superlattice period
$\mathbf{\Lambda}=m\mathbf{L}^{\mathrm{M}}_{1}+n\mathbf{L}^{\mathrm{M}}_{2}$
on the moir\'e map consists of $m$ $\alpha$-type domains and $n$ $\beta$-type domains.

In Figs.~\ref{spider_map}(b)–(d), we show similar plots for wider GNRs.
For $N \ge 40$, the web of the effective grid opens up and spans multiple AB$'$ spots across the ribbon width, corresponding to the formation of H-shaped domains.

\begin{figure*}
        \begin{center}
        \leavevmode\includegraphics[width=0.8\hsize]
        {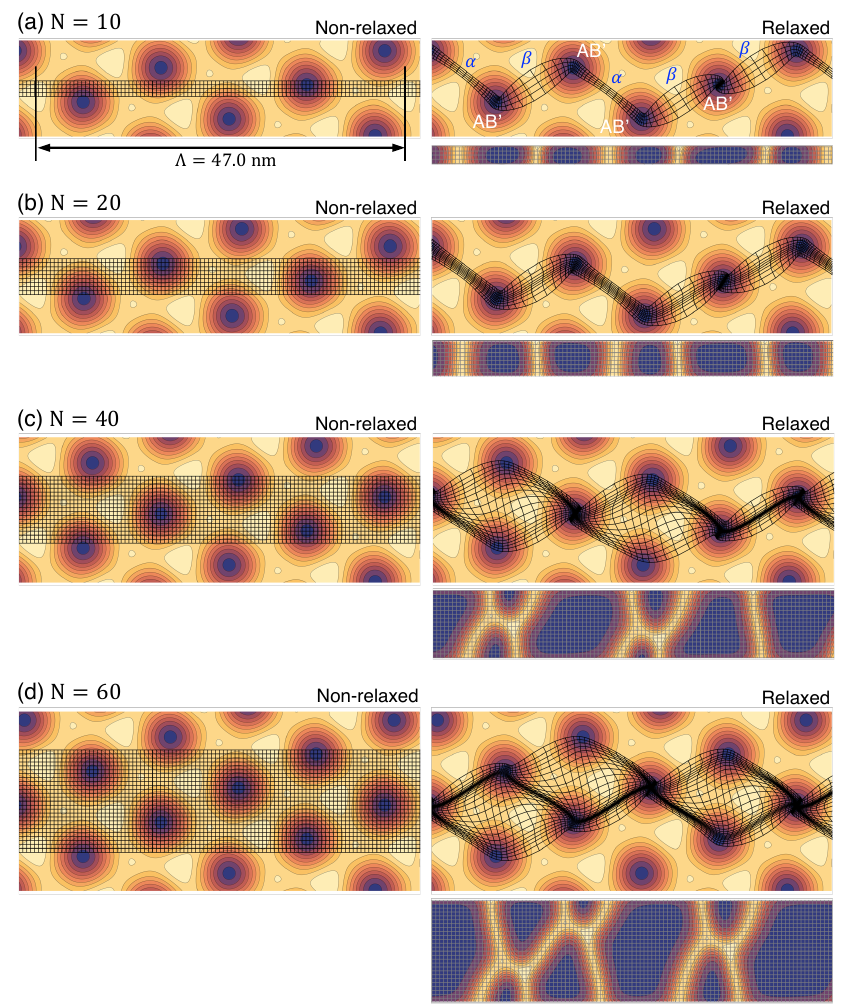}
    \caption{
Moir\'e mapping of relaxed GNR/hBN structures.
(a) $\theta_{2,3}=0.77^{\circ}$ and $N=10$: the left panel shows the non-relaxed moir\'e pattern with the effective grid of the ribbon, and the right panel shows the relaxed configuration mapped onto the same moir\'e pattern.
The lower-right inset is the moir\'e pattern on the undeformed grid [Fig.~\ref{optimized_energy_structure}(c)].
(b)–(d) Similar maps for $N=20$, $40$, and $60$.
For $N \ge 40$, the effective grid extends over multiple AB$'$ regions, forming H-shaped domains.
         }
        \label{spider_map}
        \end{center}
    \end{figure*}

\bibliography{reference.bib}

\end{document}